\PassOptionsToPackage{prologue,table,xcdraw,dvipsnames}{xcolor}
\documentclass[sigconf]{acmart}

\AtBeginDocument{%
  \providecommand\BibTeX{{%
    Bib\TeX}}}

\setcopyright{acmlicensed}
\copyrightyear{2018}
\acmYear{2018}
\acmDOI{XXXXXXX.XXXXXXX}

\acmConference[Conference acronym 'XX]{Make sure to enter the correct
  conference title from your rights confirmation emai}{June 03--05,
  2018}{Woodstock, NY}
\acmISBN{978-1-4503-XXXX-X/18/06}

\usepackage{multirow, caption, subcaption, makecell}
\usepackage{booktabs}
\usepackage[table,xcdraw,dvipsnames]{xcolor}
\usepackage{enumitem}
\usepackage{pifont}

\AtBeginDocument{%
  \providecommand\BibTeX{{%
    \normalfont B\kern-0.5em{\scshape i\kern-0.25em b}\kern-0.8em\TeX}}}
\copyrightyear{2025}
\acmYear{2025}
\setcopyright{rightsretained}
\acmConference[SIGIR '25] {Proceedings of the 48th International ACM SIGIR Conference on Research and Development in Information Retrieval}{ July 13--18, 2025}{Padua, Italy.}
\acmBooktitle{Proceedings of the 48th International ACM SIGIR Conference on Research and Development in Information Retrieval (SIGIR '25), July 13--18, 2025, Padua, Italy}
\acmISBN{979-8-4007-1592-1/25/07}
\acmDOI{10.1145/3726302.3729886}

\usepackage{etoolbox}
\makeatletter
\gdef\@copyrightpermission{
  \begin{minipage}{0.3\columnwidth}
   \href{https://creativecommons.org/licenses/by/4.0/}{\includegraphics[width=0.90\textwidth]{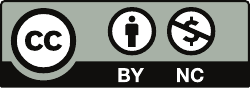}}
  \end{minipage}\hfill
  \begin{minipage}{0.7\columnwidth}
   \href{https://creativecommons.org/licenses/by/4.0/}{This work is licensed under a Creative Commons Attribution International 4.0 License.}
  \end{minipage}
  \vspace{5pt}
}
\makeatother

\begin{document}

\title{Adaptive Graph Integration for Cross-Domain Recommendation via Heterogeneous Graph Coordinators}

\author{Hengyu Zhang}
\email{hyzhang@se.cuhk.edu.hk}
\authornote{Both authors contributed equally to this research.}
\affiliation{
  \institution{\text{The Chinese University of Hong Kong}}
  \city{Hong Kong SAR}
  \country{China}}

  \author{Chunxu Shen}
  \authornotemark[1]
\email{lineshen@tencent.com}
\affiliation{
  \institution{WeChat, Tencent Inc.}
  \city{Shenzhen}
  \country{China}}

  \author{Xiangguo Sun}
  \authornote{Corresponding author.}
\email{xiangguosun@cuhk.edu.hk}
\affiliation{
  \institution{\text{The Chinese University of Hong Kong}}
  \city{Hong Kong SAR}
  \country{China}}

  \author{Jie Tan}
\email{jtan@se.cuhk.edu.hk }
\affiliation{
  \institution{\text{The Chinese University of Hong Kong}}
  \city{Hong Kong SAR}
  \country{China}}

  \author{Yu Rong}
\email{yu.rong@hotmail.com}
\affiliation{
  \institution{\text{The Chinese University of Hong Kong}}
  \city{Hong Kong SAR}
  \country{China}}

  \author{Chengzhi Piao}
\email{czpiao@comp.hkbu.edu.hk}
\affiliation{
  \institution{Hong Kong Baptist University}
  \city{Hong Kong SAR}
  \country{China}}

  \author{Hong Cheng}
\email{hcheng@se.cuhk.edu.hk }
\affiliation{
  \institution{\text{The Chinese University of Hong Kong}}
  \city{Hong Kong SAR}
  \country{China}}

  \author{Lingling Yi}
  \authornotemark[2]
\email{chrisyi@tencent.com}
\affiliation{
  \institution{WeChat, Tencent Inc.}
  \city{Shenzhen}
  \country{China}}

\renewcommand{\shortauthors}{Hengyu Zhang et al.}

\begin{abstract}

In the digital era, users typically interact with diverse items across multiple domains (e.g., e-commerce, streaming platforms, and social networks), generating intricate heterogeneous interaction graphs. Leveraging multi-domain data can improve recommendation systems by enriching user insights and mitigating data sparsity in individual domains. However, integrating such multi-domain knowledge for cross-domain recommendation remains challenging due to inherent disparities in user behavior and item characteristics and the risk of negative transfer, where irrelevant or conflicting information from the source domains adversely impacts the target domain's performance. To tackle these challenges, we propose HAGO, a novel framework with \textbf{\underline{H}}eterogeneous \textbf{\underline{A}}daptive \textbf{\underline{G}}raph co\textbf{\underline{O}}rdinators, which dynamically integrates multi-domain graphs into a cohesive structure. HAGO adaptively adjusts the connections between coordinators and multi-domain graph nodes to enhance beneficial inter-domain interactions while alleviating negative transfer. Furthermore, we introduce a universal multi-domain graph pre-training strategy alongside HAGO to collaboratively learn high-quality node representations across domains. 
Being compatible with various graph-based models and pre-training techniques, HAGO demonstrates broad applicability and effectiveness.
Extensive experiments show that our framework outperforms state-of-the-art methods in cross-domain recommendation scenarios, underscoring its potential for real-world applications. The source code is available at  \url{https://github.com/zhy99426/HAGO}.

\end{abstract}



\maketitle

\section{Introduction}

In real-world applications, users often interact with various items~\cite{li2024attention,hpmr} across multiple domains within the platform, such as different product categories in e-commerce or distinct content types in streaming services, forming complex heterogeneous interaction graphs.
However, the distribution of data across these domains is often uneven~\cite{pkef}. Some domains benefit from abundant user interaction data, while others suffer from data sparsity.
How to effectively utilize the abundant interaction data from well-established source domains to enhance recommendation performance in the target domain that faces data sparsity or holds greater strategic importance to the platform (a.k.a, cross-domain recommendation) has become an important area in academia and industry.

\begin{figure}[t]
    \centering
    \includegraphics[width=0.48\textwidth]{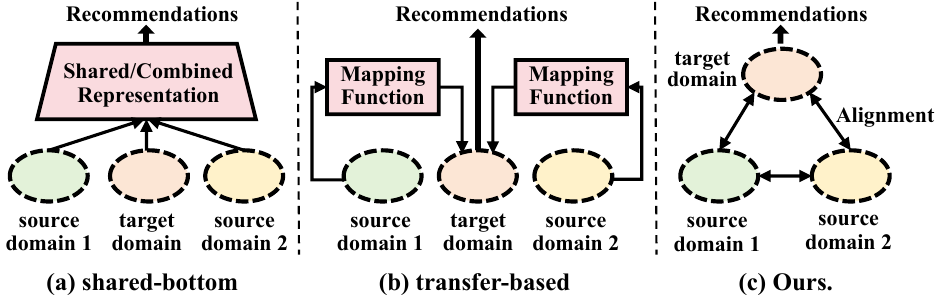}
    \caption{The schematic diagram of the existing cross-domain recommendation methods and our perspective.}
    \label{fig:intro}
\end{figure}
In recent years, various \textit{model-centric} methods have been proposed to address this problem, yet significant room for improvement remains. As illustrated in Figure \ref{fig:intro}(a), one kind of solution is to build some shared-bottom architectures by utilizing shared representations~\cite{CMF,CoNet} or representation fusion~\cite{DTCDR,NATR} to represent and transfer information across different domains. The other direction,  as shown in Figure \ref{fig:intro}(b), is to use some mapping functions or transfer layers to map user or item representations from the source domains to the target domain for information enhancement~\cite{EMCDR,CATN,SSCDR,RL-ISN}. Aligned with these two kinds of \textit{model-centric} frameworks, they also introduce some advanced training skills such as contrastive learning, to promote cross-domain transfer learning further~\cite{CUT,MOP,CCDR,MCRPL}.
Although these methods have achieved notable success, some recent theoretical studies \cite{cao2023pre} have found that the performance ceiling of model-centric methods is strictly limited by the intrinsic discrepancy among various domains, indicating that we might not further improve these methods unless we can find some data-level approaches to narrow down the natural gap among various graph domains (as presented in Figure \ref{fig:intro}(c)).

However, achieving this goal is never easy because we need to overcome at least three tough challenges:
\textbf{\textit{Challenge 1}}, designing a hand-crafted strategy to align different graph domains is far-fetched and tough. 
Interaction graph data from different domains may exhibit entirely distinct structural patterns and various types of item nodes, resulting in diverse topologies. This makes manually designed strategies far from effective and general. To this end, we need some trainable approaches to learn the potential latent data operation strategy to align these domains while preserving their important knowledge.
\textbf{\textit{Challenge 2}}, how to integrate and learn the rich knowledge from multiple source domains is tough. The complexity and diversity of these graphs not only enrich the underlying knowledge but also significantly increase the difficulty for models to capture common patterns across multiple domains.  Consequently, effectively learning useful knowledge from multi-domain source graphs presents an intricate challenge, as it is crucial for enabling models to capture cross-domain correlations among entities better.
\textbf{\textit{Challenge 3}}, how to transfer the preserved knowledge to the target is still an open problem, and currently often undergoes serious negative transfer issues. Source domains encompass a vast array of information, much of which may not be pertinent or valuable to the target domain.
Indiscriminately transferring knowledge from source domains to the target domain may adversely affect the model's performance, potentially resulting in an accuracy decline.

Inspired by the recent success of graph prompting in its powerful data operation capability~\cite{GCOPE,allinone,graphprompt}, we go beyond the previous model-centric paradigm and hope to bridge gaps between diverse graph datasets in a data-centric perspective. We offer \textbf{HAGO} framework, a data-level framework with \textbf{\underline{H}}eterogeneous \textbf{\underline{A}}daptive \textbf{\underline{G}}raph co\textbf{\underline{O}}rdinators to address the above challenges.
Specifically, to address \textbf{\textit{Challenge 1}}, we revisit the theoretical basis of graph prompts in data manipulations, and then we propose \textit{heterogeneous graph coordinators} to integrate multi-domain graphs in the pre-training stage, which shares a similar theoretical basis as graph prompts. In this way, we can get rid of hand-crafted strategies and learn a potential way to unify various source domains into a unified structure, enabling collaborative multi-domain pre-training to facilitate representation alignment among them. Notably, our graph coordinators are designed to be heterogeneous, using user/item coordinators to connect user/item nodes in graphs, thus preventing negative transfer caused by information confusion between different types of nodes.
To address \textbf{\textit{Challenge 2}}, we present the details of a \textit{multi-domain pre-training framework} with our graph coordinator mechanism. This framework is model-agnostic and can be easily used in various pre-training strategies and backbones. 
To address \textbf{\textit{Challenge 3}}, the coordinators in HAGO connect the nodes in corresponding graphs by edges with \textit{adaptive weights}, dynamically regulating knowledge transfer both within and across domains.
Moreover, we design an efficient graph prompting method that integrates the pre-trained embeddings and prompts for recommendation tasks. We conduct extensive experiments on two real-world platforms with seven different domains, results from which further demonstrate the effectiveness of our method. In summary, the main contributions of our work can be summarized as follows:
\begin{itemize}[leftmargin=*]
    \item We introduce HAGO, an effective multi-domain alignment approach with heterogeneous adaptive graph coordinators to dynamically integrate multi-domain graphs into a unified structure. HAGO enhances interconnections between multiple domains while mitigating negative transfer by adaptively adjusting the edge weights between coordinators and nodes.
    \item We develop a universal pre-training framework that can integrate various self-supervised learning algorithms alongside HAGO to collaboratively learn high-quality node representations across multiple domains.
\item We design a graph prompting approach to effectively transfer the pre-trained multi-domain knowledge to the target domain. By incorporating pre-trained embeddings with prompts, we improve recommendation accuracy in the target domain.
\item We conduct extensive experiments using seven domains on two platforms. 
Experimental results demonstrate the superior effectiveness and wide compatibility of HAGO.
\end{itemize}

\section{Background}\label{sec:back}
In this section, we introduce our motivation by comparing two perspectives of cross-domain recommendation. Then, the formal objective of this paper will be given. 

\noindent \ding{182} \textbf{Cross-domain Recommendation: A Model-centric Perspective.} A mainstream approach for cross-domain recommendation is to employ a shared-bottom architecture~\cite{CMF,CoNet,DTCDR,NATR,WCDR} and mapping functions to project knowledge from source domains to the target~\cite{EMCDR,CATN,SSCDR,RL-ISN}. These approaches are inherently \textit{model-centric}. Despite significant progress,  recent studies have revealed that model-centric approaches have an unbreakable ceiling caused by the intrinsic gap of the cross-domain data \cite{cao2023pre}. That means no matter how good the models are, their upper bound performance is always limited by the natural data gap.
Thus, it is imperative to revisit the intrinsic nature of the problem and explore ways to further break up the ceiling. 
Intuitively, narrowing the data gap among various domains could provide a higher performance improvement margin for existing models. However, traditional data dealing approaches heavily rely on manpower, making it inefficient and ineffective to manually design data operation strategies for this vision. 

\noindent \ding{183} \textbf{Our Motivation: An Attempt of Data-centric Perspective.} With the above discussion, this paper tries to explore an enhanced direction, which aims to achieve representation alignment and effective knowledge transfer from a \textit{data-centric} perspective. Unlike other linear-structured data such as images and text, graph data is highly complex and exhibits diverse non-linear structural properties. 
Nodes in a graph are not isolated; they are deeply influenced by their local neighbors. 
The connectivity among nodes serves as the carrier of information flow within the graph, reflecting the core characteristics of graph data.
Consequently, capturing the common patterns~\cite{dpn} of complex and diverse local structures across multiple graphs to align graph data at the data level particularly presents a more challenging problem. It is infeasible to leverage manually designed strategies to heal the dataset gap effectively. Fortunately, recent progress on graph prompting has demonstrated the potential effectiveness in manipulating diversified datasets to be unified~\cite{allinone,theoretical}.
Graph prompting has the potential to bridge the gaps between multi-domain pre-training and knowledge transfer.
Inspired by this, we aspire to harness the power of graph prompting to amalgamate graphs from various domains into a cohesive structure, fostering alignment and mitigating negative transfer at the data level.

\noindent \ding{184} \textbf{Formalized Objective.} We treat the given $M$ source domains data as user-item interaction graphs, denoted as $\mathcal{G}^{(k)}=(\mathcal{V}^{(k)}, \mathcal{E}^{(k)})$, $k=1, 2, \cdots, M$. $\mathcal{V}^{(k)}=\mathcal{U}^{(k)}\cup\mathcal{I}^{(k)}$, i.e., the union of the users set $\mathcal{U}^{(k)}$ and items set $\mathcal{I}^{(k)}$ in the $k$-th source domain, and $\mathcal{E}^{(k)}$ is the edge set corresponding to the adjacency matrix $\boldsymbol{A}^{(k)}\in \mathbb{R}^{\mathcal{V}^{(k)}\times \mathcal{V}^{(k)}}$ which records all the interactions between users and items in the $k$-th source domain.
The objective is to sufficiently utilize the rich information from source domains $\{\mathcal{G}^{(1)},\cdots, \mathcal{G}^{(k)},\cdots, \mathcal{G}^{(M)}\}$ to improve the Top-N recommendation performance in the target domain $\mathcal{G}^{(t)}=(\mathcal{V}^{(t)}, \mathcal{E}^{(t)})$ with the adjacency matrix $\boldsymbol{A}^{(t)}$.

\section{Method}

\begin{figure*}[!htp]
	\centering
	\includegraphics[width=0.93\textwidth]{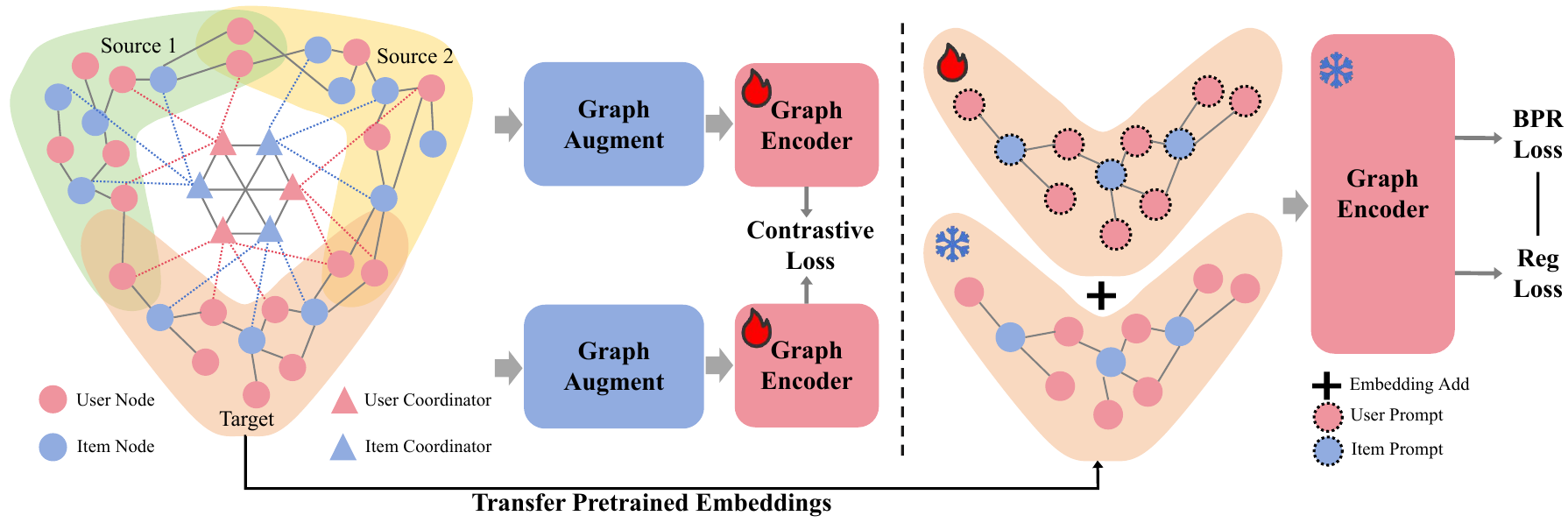}
	\caption{Illustration of the proposed HAGO, using two source domains and a target domain as examples. There is partial overlap among user nodes across different domains. For clarity, we omit the overlapping user nodes that are present in all domains. HAGO employs heterogeneous adaptive graph coordinators to organize the multi-domain pre-training collaboratively, and then leverages graph prompts to transfer the knowledge gained from pre-training to the target domain, thereby enhancing downstream recommendation tasks.}
	\label{fig:model}
\end{figure*}

\textbf{Overview:} With the above objective, we can see that we should not only try to narrow the gap between the source domains and the target domain, but also need to heal the incompatibility among multiple source domains. To this end, the
proposed HAGO is depicted in Figure \ref{fig:model}. Specifically, our framework initially employs heterogeneous adaptive graph coordinators to integrate multi-domain graphs into a unified structure and leverages self-supervised learning algorithms for collaborative multi-domain pre-training. Subsequently, graph prompting is strategically implemented to transfer the multi-domain knowledge acquired from pre-training to the target domain, aiming to improve the accuracy of recommendation.

\subsection{A Nutshell of the Backbone Model}
\label{sec:base_model}

Notably, our proposed method, HAGO, is a model-agnostic framework that is compatible with various GNN backbones. Here we briefly give a nutshell of LightGCN~\cite{LightGCN} to capture graph relations, which is widely adopted in recommendation systems. 

\noindent \ding{182} \textbf{Entity Embedding:} To capture the complex interaction patterns between users and items, we map the ID of each user or item in the given sets of users $\mathcal{U}$ and items $\mathcal{I}$ to a learnable dense vector. This operation utilizes the one-hot vector corresponding to the user or item ID to index the embedding lookup table, which can be formalized as:
\begin{equation}
    \boldsymbol{e}_u=\boldsymbol{E}^T_u \boldsymbol{o}_u, \boldsymbol{e}_i=\boldsymbol{E}^T_i \boldsymbol{o}_i,
    \label{equ:lookup}
\end{equation}
where $\boldsymbol{E}_u\in\mathbb{R}^{|\mathcal{U}|\times d}$ and $\boldsymbol{E}_i\in\mathbb{R}^{|\mathcal{I}|\times d}$ denote the user/item embedding lookup tables, and $\boldsymbol{o}_u \in \mathbb{R}^{|\mathcal{U}|}, \boldsymbol{o}_i \in \mathbb{R}^{|\mathcal{I}|}$ are the corresponding one-hot vector of user/item IDs. $|\mathcal{U}|$ and $|\mathcal{I}|$ indicate the numbers of users and items, and $d$ means the embedding size.

\noindent \ding{183} \textbf{Propagation Layer:} The graph aggregation process in LightGCN can be formalized as:
\begin{equation}
\boldsymbol{e}_u^{(l+1)}=\sum_{i \in \mathcal{N}_u} \frac{1}{\sqrt{\left|\mathcal{N}_u\right|} \sqrt{\left|\mathcal{N}_i\right|}} \boldsymbol{e}_i^{(l)}, \boldsymbol{e}_i^{(l+1)}=\sum_{u \in \mathcal{N}_i} \frac{1}{\sqrt{\left|\mathcal{N}_i\right|} \sqrt{\left|\mathcal{N}_u\right|}} \boldsymbol{e}_u^{(l)},
\end{equation}
where $\boldsymbol{e}_u^{(l)}$ and $\boldsymbol{e}_i^{(l)}$ mean the latent representations of user $u$ and item $i$ after $l$-layer graph convolutions, and $\boldsymbol{e}^{(0)}_u=\boldsymbol{e}_u$, $\boldsymbol{e}^{(0)}_i=\boldsymbol{e}_i$. $\mathcal{N}_u$ and $\mathcal{N}_i$ denote the set of items that are interacted by user $u$ and the set of users that interact with item $i$, respectively.

\subsection{Heterogeneous Adaptive Graph Coordinator}

To enhance the interconnections between multi-domains while alleviating the negative transfer phenomenon, we design effective heterogeneous adaptive graph coordinators to organize cross-domain knowledge transfer dynamically. 

\subsubsection{Heterogeneous Graph Coordinators}
The interaction data is typically modeled by constructing heterogeneous user-item bipartite graphs, which mainly include two types of nodes: users and items, as well as the user-item edges. 
Naturally, we introduce a set of user coordinators $\mathcal{C}_u^{(k)}=\{c_{u,1}^{(k)}, c_{u,2}^{(k)}, \cdots, c_{u,n}^{(k)}\}$ and item coordinators $\mathcal{C}_i^{(k)}=\{c_{i,1}^{(k)}, c_{i,2}^{(k)}, \cdots, $ $c_{i,n}^{(k)}\}$ for graph $\mathcal{G}^{(k)}$. 
Here, user coordinators are restricted to connecting only with user nodes, while item coordinators are restricted to connecting only with item nodes. 
The hyperparameter $n$ determines the number of coordinators for each type, meaning we introduce $2n$ coordinators including $n$ user coordinators and $n$ item coordinators.
Each graph coordinator connects to the nodes in the corresponding graph through dynamically weighted direct edges, and corresponds to a learnable embedding which can be obtained by retrieving the coordinator embedding table $\boldsymbol{E}_c \in \mathbb{R}^{|\mathcal{C}|\times d}$ as the same way in Equation \ref{equ:lookup}.

\subsubsection{Adaptive Connections between Coordinators and Graphs}
To enhance the effectiveness of cross-domain knowledge transfer while alleviating the negative transfer phenomenon, we propose an adaptive edge weight learning strategy to dynamically regulate the strength of information transfer between the coordinators and the nodes.
Our proposed strategy includes both soft/hard cases and can be formalized as follows:
\begin{equation}
    w=\left\{
    \begin{aligned}
    \frac{\boldsymbol{e}\cdot \boldsymbol{e}_c}{||\boldsymbol{e}||||\boldsymbol{e}_c||}, &\ \ \textit{if}\ \ \boldsymbol{e}\cdot \boldsymbol{e}_c > 0, \qquad\textit{(Soft Case)}\\
    0\qquad, &\ \ \textit{if}\ \ \boldsymbol{e}\cdot \boldsymbol{e}_c \leq 0,\qquad\textit{(Hard Case)}
    \end{aligned}\right.
\end{equation}
where $\boldsymbol{e}_c$ and $\boldsymbol{e}$ are the embeddings of the coordinator and node, and $w$ is the edge weight between them.
$(\cdot)$ means the inner product operation between two vectors.
In the soft case, the edge weights are determined by the cosine similarity between the embeddings of the coordinator and the node; in the hard case, there is no connection between the corresponding coordinator and node. Consequently, the edge weights range between 0 and 1. Following this strategy, the model can capture the degree of association between the coordinators and the nodes through embedding learning, thereby adaptively determining whether there is an edge connection and adjusting the strength of the connection.

\subsubsection{Interconnections between Coordinators}

Through the above strategy, we establish adaptive linkages between graph coordinators and their respective graphs, empowering these coordinators to bolster individual graph models by regulating the flow of information among the nodes. 
Our vision extends further, aiming to model collaborative connections across multiple domains to facilitate multi-domain representation alignment. 
Consequently, we establish direct cross-domain interconnections between the graph coordinators associated with each domain graph, effectively amalgamating the entire collection of graphs into a cohesive structure. 
These graph coordinators assume the pivotal role of information exchange hub between graphs, with the connecting edges between coordinators acting as bridges for this communication.
Notably, our proposed HAGO framework includes distinct coordinators for users and items, facilitating connections only between coordinators of different types.
In other words, user-user and item-item coordinator connections are not included. This design is based on the intuition that, even with the introduction of graph coordinators, the cohesive multi-domain graph should retain the structure of a heterogeneous user-item bipartite graph. 
This structured relationship can be depicted using an adjacency matrix $\boldsymbol{R}_{C}$:
\begin{equation}
    \begin{array}{cc}
    \boldsymbol{R}_{C}=\left[
    \begin{array}{cc}
        \boldsymbol{0}_{|\mathcal{C}_u|\times |\mathcal{C}_u|} & \boldsymbol{J}_{|\mathcal{C}_u|\times |\mathcal{C}_i|} \\
        \boldsymbol{J}_{|\mathcal{C}_i|\times |\mathcal{C}_u|} & \boldsymbol{0}_{|\mathcal{C}_i|\times |\mathcal{C}_i|}
    \end{array}\right], &
    \begin{array}{cc}
    \leftarrow user\\
    \leftarrow item
    \end{array}
    \\
    \begin{array}{cc}
    \uparrow \quad    & \qquad \uparrow\\
    user \quad & \qquad item
    \end{array} &
    \end{array}
\end{equation}
where $\boldsymbol{0}_{p\times q}$ and $\boldsymbol{J}_{p\times q}$ are zero matrix and one matrix with shape of $(p, q)$.

\subsubsection{Multi-domain Pre-training}

The graph coordinators we introduce serve as a unifying thread, weaving together the diverse domains of our dataset into an integrated tapestry through their strategic architectural integration.
Constructing upon this foundation, we introduce a universal multi-domain graph pre-training framework to collaboratively refine high-quality node representations while facilitating representation alignment. 
By integrating node-level pre-training strategies exemplified by GRACE~\cite{GRACE}, which shows strong effectiveness in learning embeddings within individual graphs through self-supervised learning, with heterogeneous adaptive graph coordinators, we achieve a globally unified multi-domain pre-training framework. 
GRACE \cite{GRACE} uses graph augmentation techniques such as edge removing and feature masking to generate pairs of positive samples, and then it employs the InfoNCE~\cite{InfoNCE} loss function to ensure that these positive samples are brought close together while negative samples are pushed apart.

In order to demonstrate the wide versatility of our framework, we subsequently adopt other state-of-the-art graph pre-training methods like BGRL \cite{BGRL} and COSTA \cite{COSTA}, integrating them into our multi-domain pre-training framework, in Section \ref{sec:compatibility}. 
COSTA builds upon GRACE by incorporating the novel Covariance-Preserving Feature Augmentation technique, which effectively tackles the bias challenges present in the original augmentation strategies.
BGRL introduces the concept of Bootstrapping Latent Loss, which is designed to align the node representations produced by the online and target encoders with different updating strategies.

\subsection{Graph Prompting for Transferring}

Prompting technology has been widely used in the NLP area, particularly in achieving general task performance of Large Language Models \cite{shu2024llm}. Inspired by this, 
graph prompting offers some prompt nodes within the graph structure or the enhancement of node features with prompt tokens, and has been widely demonstrated as a powerful method to learn various potential data operations ~\cite{GPF, GCOPE,allinone, theoretical}. This makes graph ``pre-training and prompting'' stand out in transferring knowledge for graphs compared with traditional ``pre-training and fine-tuning'' \cite{li2024graph}. Next, we introduce our designed graph prompts customized for the cross-domain recommendations.

\subsubsection{Graph Prompt}

The prompt embeddings of users and items in the target domain are integrated with the pre-trained embeddings of nodes from multi-domain pre-training to perform the recommendation task.
The input features of the model after adding graph prompts can be represented as:
\begin{equation}
    \boldsymbol{X}^\# = \{\boldsymbol{e}^*_{u_1}+\boldsymbol{p}_{u_1}, \cdots, \boldsymbol{e}^*_{u_{|\mathcal{U}|}}+\boldsymbol{p}_{u_{|\mathcal{U}|}}, \boldsymbol{e}^*_{i_1}+\boldsymbol{p}_{i_1}, \cdots, \boldsymbol{e}^*_{i_{|\mathcal{I}|}}+\boldsymbol{p}_{i_{|\mathcal{I}|}} \},
\end{equation}
where $\boldsymbol{e}^*_{u}, \boldsymbol{e}^*_{i}$ are the node embeddings of the user $u$ and item $i$ pre-trained from the multi-domain pre-training organized by HAGO, and $\boldsymbol{p}_{u}, \boldsymbol{p}_{i}$ denote the corresponding prompt embedding. These pre-trained embeddings contain a wealth of multi-domain knowledge absorbed during the pre-training phase, which can effectively enhance the model's recommendation accuracy in the target domain. 
Graph prompts serve as the bridge to align the pre-trained node embeddings to the downstream task in the target domain.

\subsubsection{Training Objective}

By utilizing the graph convolutional network as outlined in Section \ref{sec:base_model} to operate on the unified multi-domain graph framework by the graph coordinators, we derive the final node representations as follows:
\begin{equation}
    \boldsymbol{o}_u=\sum\limits_{l=0}^L \alpha_{l} \boldsymbol{e}_u^{(l)}, \boldsymbol{o}_i=\sum\limits_{l=0}^L \alpha_{l} \boldsymbol{e}_i^{(l)},
\end{equation}
where $\boldsymbol{o}_u, \boldsymbol{o}_i$ denote the final representations of user $u$ and item $i$. $\alpha_{l}$ is the weight of the representation after $l$-layers which satisfies $\sum\limits_{l=0}^L \alpha_{l}=1$, where $L$ is the number of layers in the model.

We leverage the Bayesian Personalized Ranking (BPR) loss function~\cite{BPR}, a prevalent choice for pairwise ranking within the recommendation research area, to train our model's performance on the target domain's recommendation tasks. 
The BPR loss leads the model to assign higher predictive scores to user-item interactions that have occurred compared to those that have not. Thus, the final training objective can be articulated as follows:
\begin{equation}
    \begin{aligned}
    \mathcal{L} &= \mathcal{L}_{BPR} + \mathcal{L}_{Reg}
    \\&=-\frac{1}{|\mathcal{B}|}\sum\limits_{(u,i,j)\in \mathcal{B}}\mathrm{ln} \sigma(\hat y_{ui}-\hat y_{uj}) + \lambda ||\Theta||^2,
    \end{aligned}
\end{equation}
where the second term is the $L2$-regularization to prevent overfitting. $(u,i,j)$ from batch $\mathcal{B}$ is a triple including user $u$ and item $i,j$.
The interaction between user $u$ and item $i$ represents a positive pair, while the absence of interaction between user $u$ and item $j$ signifies a negative pair.
$\hat y_{ui} = \boldsymbol{o}_u \cdot \boldsymbol{o}_i$ and $\hat y_{uj} = \boldsymbol{o}_u \cdot \boldsymbol{o}_j$ are the predicted value of interaction occurring between user $u$ and item $i,j$. $\Theta$ denotes the entire ensemble of trainable parameters that the model employs and $\lambda$ controls the regularization degree.

\subsection{Why It Works?}

\noindent \ding{182} \textbf{Theoretical Support:} Our proposed framework shares a common theoretical bedrock with graph prompts designed for knowledge transfer. We draw inspiration from the ``Prompt as Graph" paradigm \cite{Prog,graphprompt}, which introduces multiple additional prompt nodes with the inner structure, which correspond to learnable embeddings, into the original graph and connects them through specific structures \cite{allinone,PRODIGY}. The primary difference is that while graph prompts are generally targeted at downstream tasks, we utilize the term ``graph coordinators" for the pre-training phase. 
Previous studies \cite{GPF,allinone} have demonstrated that graph prompts can emulate any graph operations, such as node/edge addition and deletion.
Recent studies~\cite{theoretical} have shown that for a given graph $\mathcal{G}$, we can find some potential graph transformation $T(\cdot)$ to translate the original graph as a new one like $T(\mathcal{G})$ such that the pre-trained graph encoder $f_{\theta^*}(\cdot)$ can approximate to an optimal function $C$ for the target domain task, which can be described as:
\begin{equation}
    f_{\theta^*}(T(\mathcal{G})) = C(\mathcal{G}).
\end{equation}
In other words, graph prompts can be treated as a learnable method to find such data transformation~\cite{GPF, allinone} to approximate optimal in the target task~\cite{theoretical}. Specifically, given a pre-trained graph encoder $f_{\theta^*}(\cdot)$ with fixed parameters $\theta^*$ and learnable graph prompt $P(\cdot)$, it can be shown that:
\begin{equation}
f_{\theta^*}(P(\mathcal{G}))=f_{\theta^*}(T(\mathcal{G}))+O_{fp},
\end{equation}
where $O_{fp}$ is the bounded error associated with $f_{\theta^*}(\cdot)$ and $P(\cdot)$.
This implies that graph prompts have the capability to be equivalent to certain graph transformation operations that enable the transformed graph, once processed by the pre-trained graph encoder, to produce optimal embeddings. In this paper, our heterogeneous adaptive graph coordinators can be treated as a variant of graph prompts, which are used in the pre-training stage, and thus share the same foundation for data operations. Based on this idea, we give a data-centric solution to heal the gap among multiple source domains and the target domain such that the performance ceiling of knowledge transfer can be further improved as mentioned in our motivation (Section \ref{sec:back}). 

\noindent \ding{183} \textbf{Compatibility Analysis:} Our proposed HAGO is model-agnostic because it extends the graph structure without regarding to message passing and aggregation within the graph, making it compatible with various graph-based backbone networks. 
It can also be seamlessly integrated with various graph pre-training algorithms. 
In our subsequent experiments, comprehensive compatibility analysis demonstrates the broad compatibility and wide effectiveness of HAGO with different backbones and pre-training methods.

\noindent \ding{184} \textbf{Complexity Analysis:} The time complexity of our proposed HAGO framework is intrinsically linked to that of the backbone model. The additional time complexity introduced by HAGO primarily stems from the heterogeneous adaptive graph coordinators and their connections with the multi-domain graphs. Our backbone model, LightGCN, has a time complexity of $\mathcal{O}(|\mathcal{E}^{(t)}|Ld)$, where $|\mathcal{E}^{(t)}|$ represents the number of edges in the target-domain graph, and $L$ and $d$ denote the number of layers and the dimension of the network, respectively. Consequently, for HAGO, the time complexity during the multi-domain pre-training phase is $\mathcal{O}((\sum\limits_{k=1}^M|\mathcal{E}^{(k)}|+|\mathcal{E}^{(t)}|+|\mathcal{E}_C|)Ld)$ where $|\mathcal{E}_C|$ is the number of edges introduced by HAGO; and when transferring to downstream recommendation tasks via graph prompts, the time complexity remains $\mathcal{O}(|\mathcal{E}^{(t)}|Ld)$.

\begin{table*}[!htbp]
\caption{Overall comparison results of our proposed HAGO and state-of-the-art baselines on two real-world multi-domain datasets. 
The best-performing method is highlighted in bold, while the second-best is indicated with an underline. An asterisk (*) denotes statistical significance (p < 0.05) when comparing HAGO to the best-performing baseline results.}

\begin{tabular}{l|l|ccc|ccccccc}
\toprule
                          &                                  & \multicolumn{3}{c|}{Single-domain Method} & \multicolumn{7}{c}{Cross-domain Method}                       \\ \cline{3-12} 
\multirow{-2}{*}{Dataset} & \multirow{-2}{*}{Metric}         & NGCF        & LightGCN      & XSimGCL     & CMF    & EMCDR  & SSCDR  & DeepAPF & BiTGCF & CUT    & Ours.  \\ \hline
                          & Recall@10 & 0.1195      & 0.1665        & 0.1660      & 0.1143 & 0.1264 & 0.1430 & 0.1442  & 0.1543 & \underline{0.1669} & \textbf{0.1689}$^*$ \\
                          & HR@10     & 0.2382      & 0.3116        & 0.3117      & 0.2138 & 0.2498 & 0.2801 & 0.2820  & 0.2879 & \underline{0.3123} & \textbf{0.3168}$^*$ \\
                          & NDCG@10   & 0.0807      & 0.1210        & \underline{0.1213}      & 0.0728 & 0.0870 & 0.1024 & 0.1029  & 0.1103 & 0.1211 & \textbf{0.1227}$^*$ \\
\multirow{-4}{*}{Douban}  & MRR                              & 0.0997      & 0.1516        & 0.1526      & 0.0842 & 0.1079 & 0.1290 & 0.1288  & 0.1362 & \underline{0.1532} & \textbf{0.1535}$^*$ \\ \hline
                          & Recall@10 & 0.1064      & 0.1627        &  0.1630     & 0.1195 & 0.1236 & 0.1481 & 0.1542  & 0.1605 & \underline{0.1641} & \textbf{0.1662}$^*$ \\
                          & HR@10    & 0.1236      & 0.1902        &  0.1913     & 0.1400 & 0.1399 & 0.1724 & 0.1796  & 0.1874 & \underline{0.1918} & \textbf{0.1950}$^*$ \\
                          & NDCG@10   & 0.0675      & 0.1141        &  0.1147     & 0.0667 & 0.0834 & 0.1030 & 0.1051  & 0.1131 & \underline{0.1158} & \textbf{0.1177}$^*$ \\
\multirow{-4}{*}{Amazon}  & MRR                              & 0.0601      & 0.1074        & \underline{0.1088}      & 0.0542 & 0.0752 & 0.0951 & 0.0963  & 0.1056 & 0.1055 & \textbf{0.1116}$^*$ \\ \bottomrule
\end{tabular}
\label{tab:overall}
\end{table*}

\section{Evaluation Analysis}

\subsection{Experiment Settings}

\begin{table}[!ht]

\caption{Dataset Statistics}
\resizebox{0.48\textwidth}{!}{
\centering
\begin{tabular}{llcccc}
\toprule
dataset & domain & \#users & \#items & \#interactions & sparsity \\ \midrule
\multirow{3}{*}{Douban}      & Music                      & 15,996  & 39,749  & 1,116,984      & 99.82\%  \\
                             & Movie                      & 22,041  & 25,802  & 2,552,305      & 99.55\%  \\ 
                             & Book                       & 18,086  & 33,068  & 809,248        & 99.86\%  \\\midrule
\multirow{4}{*}{Amazon}      & Book                      & 135,110 & 115,173 & 4,042,382      & 99.97\%  \\

                             & Electronic                & 112,988 & 41,905  & 1,880,865      & 99.96\%  \\
                             & Movie                     & 26,969  & 18,564  & 762,957        & 99.85\%  \\ 
                             & Toy                       & 11,687  & 8,289   & 202,230        & 99.79\%  \\\bottomrule

\end{tabular}
}
\label{tab:dataset}
\end{table}

\noindent \ding{182} \textbf{Data Preparation and Setup:} We conduct comprehensive experiments on two popular public datasets: \textbf{Amazon dataset \cite{Amazon}\footnote{http://jmcauley.ucsd.edu/data/amazon}} is gathered from relevant product evaluations on Amazon.com. For our study, we have chosen four categories: ``\textit{Books}'', ``\textit{Electronics}'', ``\textit{Movies and TV}'', and ``\textit{Toys and Games}''; \textbf{Douban dataset \cite{Recbole}\footnote{https://recbole.s3-accelerate.amazonaws.com/CrossDomain/Douban.zip}} is collected from Douban website, which is a large-scale platform for rating and reviewing movies, TV shows, books, and music. This dataset includes three domain datasets: ``\textit{Douban-Movie}'', ``\textit{Douban-Book}'', and ``\textit{Douban-Music}''. In our experiments, we adopt the domain with the least interactions in each dataset as the target domain, namely ``\textbf{\textit{Douban-Book}}'' and ``\textbf{\textit{Amazon-Toy}}'', with the other domains serving as the source domains.
More detailed statistics of the datasets are given in Table \ref{tab:dataset}.

\noindent \ding{183} \textbf{Evaluation Metrics:} We assess the performance of our method by four widely used metrics in recommendations: Recall@10,  HR@10 (Hit Ratio), NDCG@10 (Normalized Discounted Cumulative Gain), and MRR (Mean Reciprocal Rank) for top-10 recommendation. Kindly note that we fully rank the entire items within the datasets for calculating the above top-10 recommendation performance rather than evaluating these metrics by pre-sampling a small group of the total items. This is closer to the real industrial applications and raises more challenging tasks for the compared models.

\noindent \ding{184} \textbf{Baselines:} We compare our method with \textbf{NINE} state-of-the-art (SOTA) methods, which can be grouped into two categories: (1)  Single-domain Recommendation Methods with \textbf{NGCF}~\cite{NGCF}, \textbf{LightGCN}~\cite{LightGCN}, and \textbf{XSimGCL}~\cite{XSimGCL}. (2) Cross-domain Recommendation Methods with \textbf{CMF}~\cite{CMF}, \textbf{EMCDR}~\cite{EMCDR}, \textbf{DeepAPF}~\cite{DeepAPF}, \textbf{SSCDR}~\cite{SSCDR}, \textbf{BiTGCF}~\cite{BiTGCF}, and \textbf{CUT}~\cite{CUT}.

\noindent \ding{185} \textbf{Implementation Details:} All the baselines and our proposed HAGO are implemented based on the PyTorch~\cite{pytorch}. Specifically, CUT~\cite{CUT} is implemented following its official code, and other baselines are developed based on the open-source RecBole library~\cite{Recbole}. For a fair comparison, we set the embedding size to $64$ for all methods. 
For all methods that use LightGCN~\cite{LightGCN} as the backbone, we set the number of layers to $2$ and the weight $\boldsymbol{\alpha}$ of each layer's representation to $[0, 0, 1]$, i.e., LightGCN-single in LightGCN paper~\cite{LightGCN}.
We conduct a grid search for the key hyperparameters for all baselines. For our proposed HAGO, we perform pretraining using Adam optimizer with a learning rate of $0.005$ and batch size of $1,024$, while a learning rate of $0.1$ and batch size of $4,096$ for training. And the numbers of user and item graph coordinators are both set to $n=5$.

\subsection{Overall Performance}

The overall performance comparison of our proposed HAGO and state-of-the-art baselines are shown in Table \ref{tab:overall}.
According to the experiment results, we have  the following observations:

\begin{itemize}[leftmargin=*]
    \item The approaches with LightGCN as the backbone network (i.e., XSimGCL, BiTGCF, CUT, HAGO) demonstrate exceptional performance, which benefits from LightGCN's strong generalization capabilities and its efficient capturing of intricate, granular collaborative signals. Compared to other backbones, such as Matrix Factorization (MF), LightGCN employs a simple yet effective approach to extract higher-order connectivity from the user-item interaction graph.

    \item Cross-domain knowledge transfer does not always result in superior performance.
    The performance of many cross-domain recommendation methods often falls short when compared to the elite single-domain counterparts.
    XSimGCL demonstrates superior performance over the majority of its cross-domain peers on both two datasets. This might be caused by that the performance gains derived from cross-domain knowledge transfer pale compared to the substantial advancements achieved by improving the model's underlying architecture, and the knowledge transfer from the source domain to the target domain in many cross-domain methods may not be effective enough, and it might even result in the negative transfer issue.
    
    \item Our proposed HAGO consistently achieves the best performance across all four metrics on two multi-domain datasets, demonstrating its superior effectiveness.
    HAGO owes its success to employing graph coordinators that seamlessly integrate multi-domain graphs into a cohesive framework for multi-domain collaborative pre-training. These graph coordinators adaptively connect with corresponding domain graphs, dynamically adjusting the information communication within and across domains, achieving information enhancement effectively while mitigating negative transfer. Moreover, it adopts graph prompts to efficiently transfer valuable knowledge from multiple source domains to the target domain, thereby boosting the recommendation performance.

\end{itemize}

\subsection{Effectiveness Analysis on Healing Negative Transfer Problem}

To delve into the effect of the graph coordinators in our proposed HAGO framework and to evaluate its capability to mitigate negative transfer, we craft a variety of model variants predicated on distinct coordinator designs:
\begin{itemize}[leftmargin=*]
    \item \textbf{w/o GO}: Without Graph Coordinator implies that the graph pre-training process is conducted in an isolated manner across the various domain graphs.
    \item \textbf{HomoGO}: This design features homogeneous graph coordinators that establish direct connections, whose edge weight is 1, with all nodes within the associated domain graph, treating user and item nodes uniformly.
    \item \textbf{HeterGO}: This design employs heterogeneous graph coordinators that connect exclusively with all user nodes through the user coordinators and with all item nodes through the item coordinators within the respective domain graph.
\end{itemize}

\begin{table}[!htp]
\caption{Comparative analysis results of different Graph Coordinator strategies. {\color{RedOrange}Orange} and {\color{BlueGreen}cyan} indicate performance gains or losses compared to the Backbone, with deeper shades representing greater relative changes.}
\begin{tabular}{lcccc} \toprule
\multicolumn{5}{c}{Douban}                         \\ \hline
Method     & Recall@10 & HR@10  & NDCG@10 & MRR    \\ \hline
Backbone       & 0.1665                                 & 0.3116                             & 0.1210                               & 0.1516 \\
w/o GO  &  \cellcolor{RedOrange!6.5}{0.1671}    & \cellcolor{RedOrange!2.9}{0.3121} & \cellcolor{RedOrange!5.9}{0.1214}  & \cellcolor{RedOrange!2.4}{0.1518} \\
HomoGO  & \cellcolor{RedOrange!2.2}{0.1667}    & \cellcolor{BlueGreen!6}{0.3115} & 0.1210  & \cellcolor{BlueGreen!11}{0.1515} \\
HeterGO & \cellcolor{RedOrange!11.9}{0.1676}    & \cellcolor{RedOrange!8.1}{0.3130} & \cellcolor{RedOrange!10.4}{0.1217}  & \cellcolor{RedOrange!3.6}{0.1519} \\
HAGO & \cellcolor{RedOrange!25.9}{0.1689}    & \cellcolor{RedOrange!30}{0.3168} & \cellcolor{RedOrange!25.3}{0.1227}   & \cellcolor{RedOrange!22.5}{0.1535} \\ \bottomrule
\multicolumn{5}{c}{Amazon}                         \\ \hline
Method     & Recall@10 & HR@10  & NDCG@10 & MRR    \\ \hline
Backbone       & 0.1627                                 & 0.1902                             & 0.1141                               & 0.1074 \\
w/o GO  & \cellcolor{BlueGreen!14}{0.1624}    & \cellcolor{BlueGreen!4}{0.1901} & \cellcolor{BlueGreen!13}{0.1139}  & \cellcolor{BlueGreen!25}{0.1067} \\
HomoGO  & \cellcolor{RedOrange!1.4}{0.1630}    & \cellcolor{BlueGreen!4}{0.1901} & \cellcolor{RedOrange!20.2}{0.1171}  & \cellcolor{RedOrange!5.7}{0.1082} \\
HeterGO & \cellcolor{RedOrange!9}{0.1646}    & \cellcolor{RedOrange!6.1}{0.1917} & \cellcolor{RedOrange!23.5}{0.1176}  & \cellcolor{RedOrange!25.7}{0.1110} \\
HAGO & \cellcolor{RedOrange!16.5}{0.1662}    & \cellcolor{RedOrange!19.4}{0.1950} & \cellcolor{RedOrange!24.2}{0.1177}  & \cellcolor{RedOrange!30}{0.1116} \\ \bottomrule
\end{tabular}
\label{tab:ablation}
\end{table}

The comparative experimental results of different model variants on two datasets are presented in Table \ref{tab:ablation}.
For a fair comparison, the number of coordinators is set to be consistent for HomoGO, HeterGO, and HAGO. 
From the experimental outcomes, we can draw the following observations:

\begin{itemize}[leftmargin=*]
    \item We can observe negative transfer phenomena happen in some cases, as indicated by the {\color{BlueGreen}cyan}-shading in the table. In the Amazon dataset, the performance of w/o GO declines compared to the Backbone Model; in the Douban dataset, HomoGO, which employs homogeneous graph coordinators for cross-domain information transfer during the pre-training phase, also experiences a certain degree of performance damage relative to w/o GO.
    In fact, the potential for HomoGO to lead to negative transfer effects is foreseeable.
    Homogeneous coordinators in HomoGO that indiscriminately connect to user and item nodes in a heterogeneous interaction graph can lead to information confusion across different types of nodes.
    Hence, we employ a heterogeneous design for the graph coordinators in HAGO to avoid this conflict.

    \item  HAGO consistently leads to superior recommendation performance over other model variants across both datasets, securing significant improvements.
    The success of HAGO can be primarily attributed to two key aspects.
    Firstly, HAGO integrates multi-domain graphs into a cohesive structure via heterogeneous graph coordinators,  effectively avoiding the potential negative transfer caused by information confusion between different types of nodes as seen in HomoGO.
    This is also substantiated by the consistent improvements in HeterGO results compared to HomoGO.
    Secondly, HAGO addresses the negative transfer problem effectively by employing an adaptive, trainable connection architecture that links the graph coordinators with multi-domain graphs.
    This connection structure facilitates dynamic adjustment of the transfer of more valuable information to the target domain.
    
\end{itemize}

\subsection{Compatibility Analysis with Different Backbones and Pre-training Strategies}
\label{sec:compatibility}

Our proposed HAGO is not limited to a specific model but is designed as a flexible architecture that can be effortlessly combined with a variety of backbone networks serving as graph encoders, as well as with diverse graph pre-training algorithms. To demonstrate the adaptability of HAGO, we conduct comparative experiments by integrating it with other notable graph neural networks, including GCN \cite{GCN} and GAT \cite{GAT}. Furthermore, we investigate the incorporation of other self-supervised learning strategies for pre-training the graph, such as BGRL \cite{BGRL} and COSTA \cite{COSTA}.
\begin{table}[!htp]
\caption{Compatibility Analysis with Different Backbones. {\color{RedOrange}Orange} and {\color{BlueGreen}cyan} indicate performance gains or losses compared to the corresponding Backbone, with deeper shades representing greater relative changes.}
\resizebox{0.48\textwidth}{!}{
\begin{tabular}{lcccc}
\toprule
\multicolumn{5}{c}{Douban}                               \\ \hline
Method           & Recall@10 & HR@10  & NDCG@10 & MRR    \\ \hline
GCN              & 0.1198    & 0.2392 & 0.0819  & 0.1023 \\
HAGO(GCN)      & \cellcolor{RedOrange!18}{0.1250}    & \cellcolor{RedOrange!9}{0.2446} & \cellcolor{RedOrange!19}{0.0855}  & \cellcolor{RedOrange!14}{0.1056} \\\hline
GAT              & 0.1178    & 0.2315 & 0.0795  & 0.0977 \\
HAGO(GAT)      & \cellcolor{RedOrange!25}{0.1248}    & \cellcolor{RedOrange!18}{0.2416} & \cellcolor{RedOrange!30}{0.0850}  & \cellcolor{RedOrange!29}{0.1043} \\ \bottomrule
\multicolumn{5}{c}{Amazon}                               \\ \hline
Method           & Recall@10 & HR@10  & NDCG@10 & MRR    \\ \hline
GCN              & 0.1183    & 0.1444 & 0.0802  & 0.0682 \\
HAGO(GCN)      & \cellcolor{RedOrange!26}{0.1274}    & \cellcolor{RedOrange!14}{0.1506} & \cellcolor{RedOrange!7}{0.0820}  & \cellcolor{RedOrange!29}{0.0742} \\\hline
GAT              & 0.1316    & 0.1539 & 0.0929  & 0.0878 \\
HAGO(GAT)      & \cellcolor{RedOrange!9}{0.1350}    & \cellcolor{RedOrange!8}{0.1571} & \cellcolor{RedOrange!5}{0.0939}  & \cellcolor{BlueGreen!5}{0.0874} \\ \bottomrule
\end{tabular}}
\label{tab:backbone}
\end{table}

\begin{table}[!htp]
\caption{Compatibility Analysis with Different Graph Pretraining Methods. {\color{RedOrange}Orange} and {\color{BlueGreen}cyan} indicate performance gains or losses compared to the Backbone, with deeper shades representing greater relative changes.}
\resizebox{0.48\textwidth}{!}{
\begin{tabular}{lcccc}
\toprule
\multicolumn{5}{c}{Douban}                                                                                                                     \\ \hline
Method           & Recall@10 &  HR@10 & NDCG@10 & MRR    \\ \hline
Backbone       & 0.1665                                 & 0.3116                             & 0.1210                               & 0.1516 \\ \hline
Backbone+BGRL  & \cellcolor{RedOrange!10}{0.1679}                                 & \cellcolor{RedOrange!12}{0.3145}                             & \cellcolor{RedOrange!8}{0.1218}                               & \cellcolor{RedOrange!1}{0.1517} \\
HAGO(BGRL)       & \cellcolor{RedOrange!30}{0.1704}                                 & \cellcolor{RedOrange!22}{0.3166}                             & \cellcolor{RedOrange!25}{0.1234}                               & \cellcolor{RedOrange!18}{0.1537} \\ \hline
Backbone+COSTA & \cellcolor{RedOrange!11}{0.1679}                                 & \cellcolor{RedOrange!10}{0.3141}                             & \cellcolor{RedOrange!3}{0.1213}                               & \cellcolor{BlueGreen!4}{0.1511} \\
HAGO(COSTA)      & \cellcolor{RedOrange!25}{0.1698}                                 & \cellcolor{RedOrange!11}{0.3143}                             & \cellcolor{RedOrange!22}{0.1231}                               & \cellcolor{RedOrange!8}{0.1523} \\ \bottomrule
\multicolumn{5}{c}{Amazon}                                                                                                                     \\ \hline
Method           &  Recall@10 & HR@10 & NDCG@10 & MRR    \\ \hline
Backbone       & 0.1627                                 & 0.1902                             & 0.1141                               & 0.1074 \\ \hline
Backbone+BGRL  & \cellcolor{RedOrange!4}{0.1642}                                 & \cellcolor{RedOrange!1}{0.1904}                             & \cellcolor{RedOrange!9}{0.1164}                               & \cellcolor{RedOrange!12}{0.1104} \\
HAGO(BGRL)       & \cellcolor{RedOrange!14}{0.1676}                                 & \cellcolor{RedOrange!12}{0.1953}                             & \cellcolor{RedOrange!23}{0.1199}                               & \cellcolor{RedOrange!30}{0.1144} \\ \hline
Backbone+COSTA & \cellcolor{RedOrange!4}{0.1640}                                 & \cellcolor{RedOrange!4}{0.1918}                             & \cellcolor{RedOrange!8}{0.1160}                               & \cellcolor{RedOrange!12}{0.1104} \\
HAGO(COSTA)      & \cellcolor{RedOrange!15}{0.1679}                                 & \cellcolor{RedOrange!13}{0.1955}                             & \cellcolor{RedOrange!16}{0.1181}                               & \cellcolor{RedOrange!19}{0.1119} \\ \bottomrule
\end{tabular}}
\label{tab:pretrain}
\end{table}

The compatibility analysis experiments of HAGO combined with different backbones and graph pre-training algorithms are presented in Table \ref{tab:backbone} and \ref{tab:pretrain}, respectively. From the experimental results, we can see
 that HAGO consistently enhances the performance of all backbones, and achieves unified modeling that improves the performance of graph pre-training algorithms on isolated graphs. These results demonstrate the broad compatibility and universal applicability of the HAGO framework we proposed.

\subsection{Impact Analysis of Coordinator Number}

\begin{figure}[t]
\centering
\begin{subfigure}{0.49\linewidth}
    \centering
    \includegraphics[width=0.96\linewidth]{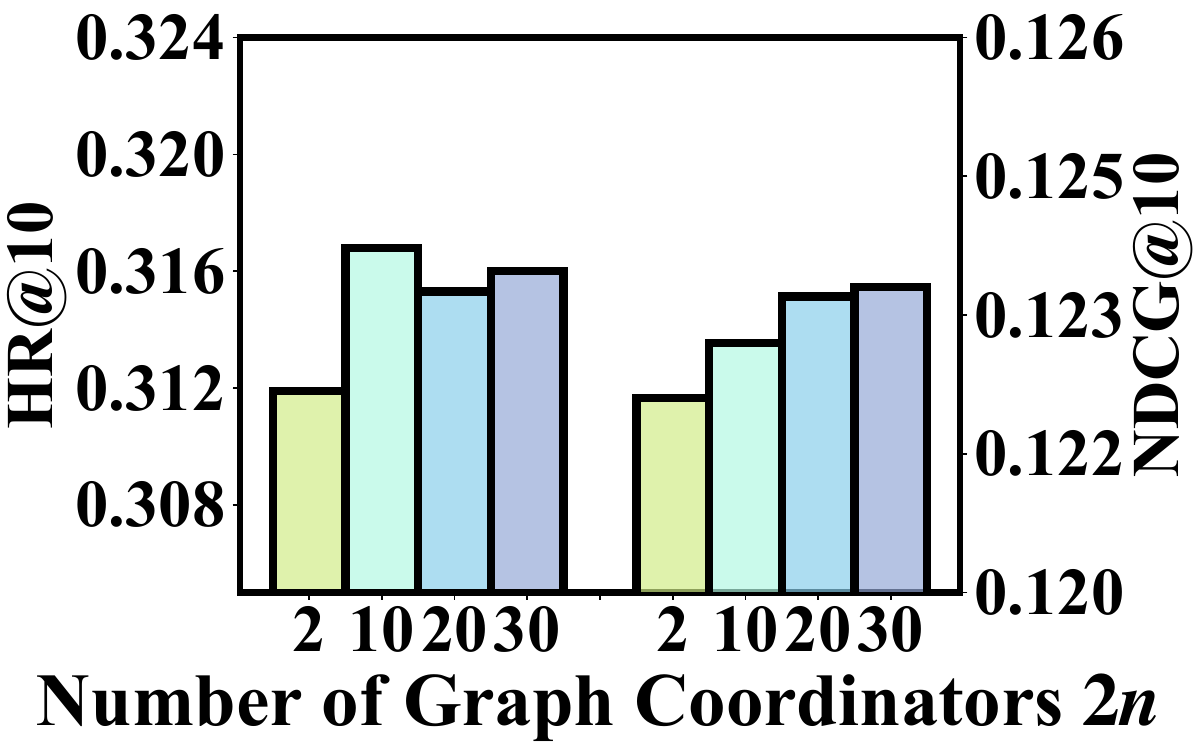}
    \caption{Douban}
\end{subfigure}
\begin{subfigure}{0.49\linewidth}
    \centering
    \includegraphics[width=0.96\linewidth]{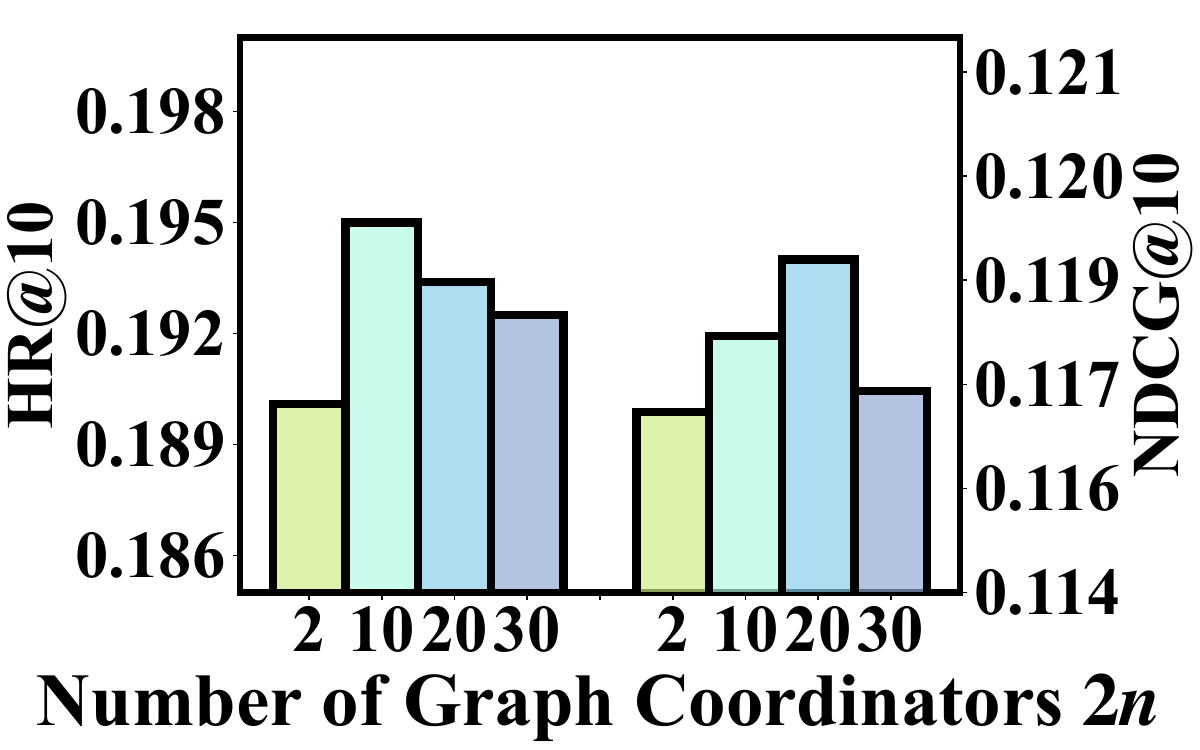} 
    \caption{Amazon}
\end{subfigure}
    \caption{Performances with different numbers of graph coordinators $2n$.}
\label{fig:hyper}
\end{figure}

We conduct comparative experiments around HAGO with different numbers of graph coordinators, and the results of HR@10 and NDCG@10 are presented in Figure \ref{fig:hyper}. 
Notably, HAGO includes both user and item graph coordinators, and the number of each type of coordinator is $n$.
In experiments, we simultaneously vary both the numbers of user and item coordinators.
Thus, the total number of graph coordinators in HAGO is $2n$.
While examining the impact of the number of coordinators, we keep other hyperparameters fixed.

Observing the results, it is evident that the optimal performance of HAGO is reached when the number of coordinators is moderate. When there are too few coordinators, they are inadequate to handle the demanding task of transferring knowledge both within and across domains efficiently. As the number of coordinators increases, the marginal benefits to model performance from the additional coordinators diminish, and there may even be a decline in performance, potentially due to excessive knowledge transfer.

\subsection{Visualization Analysis}
\label{sec:visualization}

To investigate whether the proposed HAGO framework can effectively align the representation spaces and facilitate cross-domain knowledge transfer among multiple domains, we conduct a visualization analysis of the pre-trained embeddings for item nodes in each domain, following the analytical methods in \cite{SimGCL}. We randomly sample 2000 items from each of the three domains in the Douban dataset (Douban-Music, Douban-Movie, and Douban-Book). Subsequently, we utilize t-SNE~\cite{t-sne} to perform dimensionality reduction to map these item embeddings to two-dimensional feature vectors in the $\mathbb{R}^2$ and normalize them onto the unit circle. Employing Gaussian kernel estimation~\cite{kde}, we plot the feature distribution and corresponding angle distribution of the item samples from each domain as they are mapped onto the unit circle.

\begin{figure}[!h]
\centering
\begin{subfigure}{\linewidth}
\begin{minipage}[t]{0.328\linewidth}
    \centering
    \includegraphics[width=\linewidth]{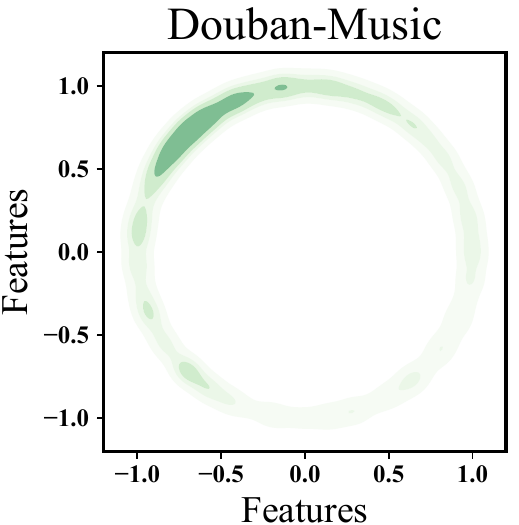}
    \includegraphics[width=\linewidth]{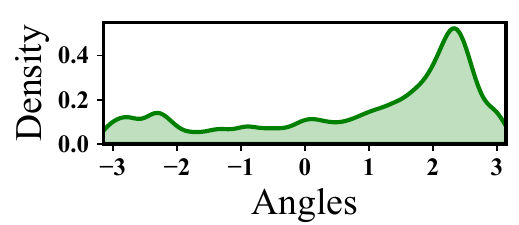}
\end{minipage}
\begin{minipage}[t]{0.328\linewidth}
    \centering
    \includegraphics[width=\linewidth]{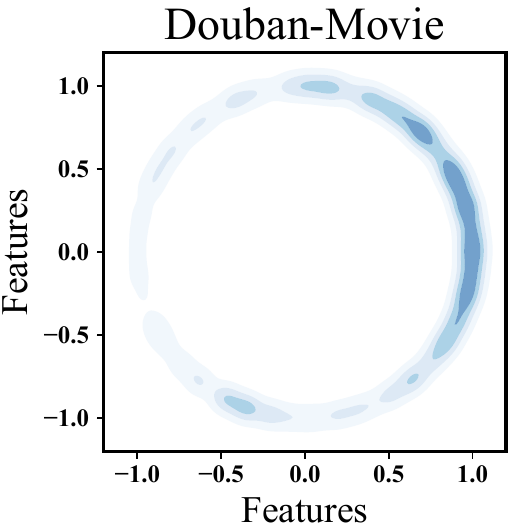} 
    \includegraphics[width=\linewidth]{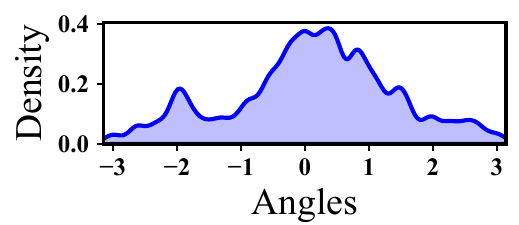}
\end{minipage}
\begin{minipage}[t]{0.328\linewidth}
    \centering
    \includegraphics[width=\linewidth]{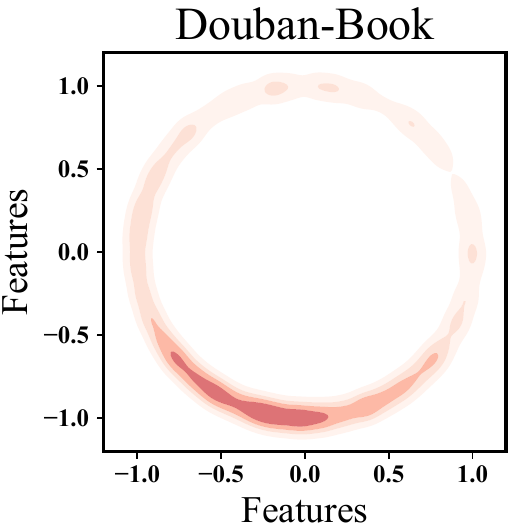}
    \includegraphics[width=\linewidth]{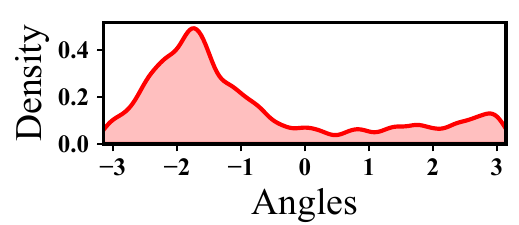}
\end{minipage}
\caption{w/o GO}
\end{subfigure}
\begin{subfigure}{\linewidth}
\begin{minipage}[t]{0.328\linewidth}
    \centering
    \includegraphics[width=\linewidth]{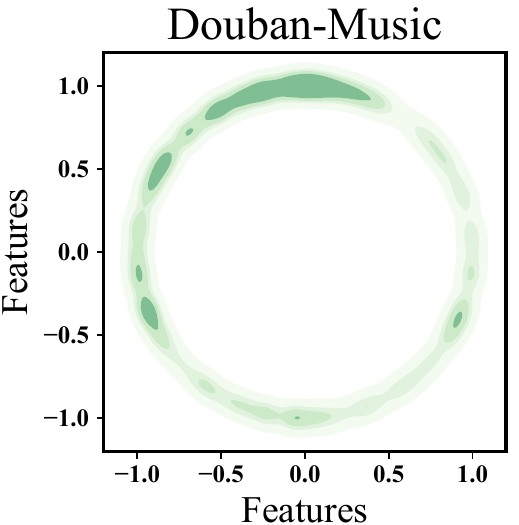}
    \includegraphics[width=\linewidth]{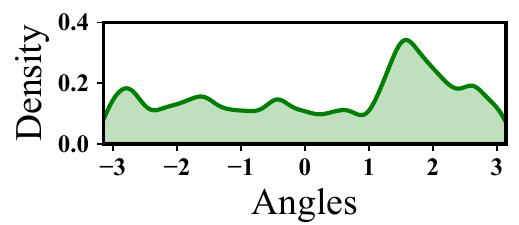}
\end{minipage}
\begin{minipage}[t]{0.328\linewidth}
    \centering
    \includegraphics[width=\linewidth]{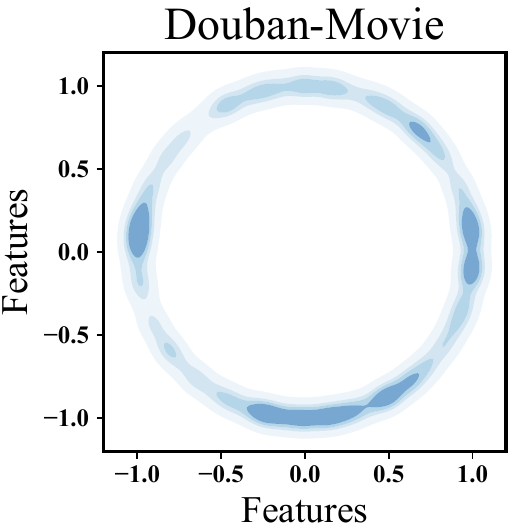} 
    \includegraphics[width=\linewidth]{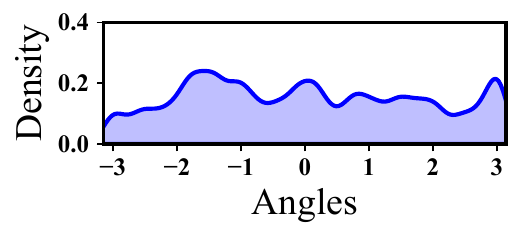}
\end{minipage}
\begin{minipage}[t]{0.328\linewidth}
    \centering
    \includegraphics[width=\linewidth]{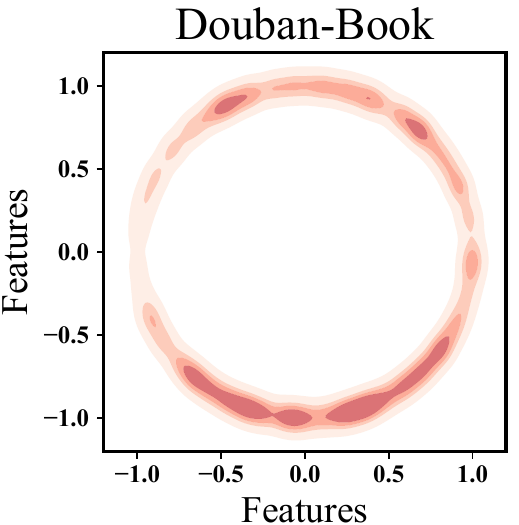}
    \includegraphics[width=\linewidth]{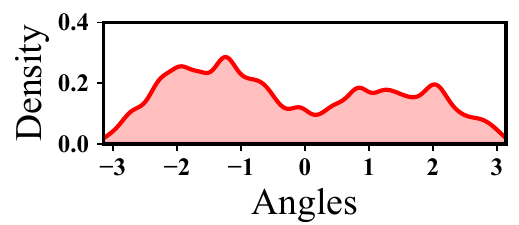}
\end{minipage}
\caption{HAGO}
\end{subfigure}
    \caption{Visualization Analysis of Item Embeddings in Douban Dataset. We perform dimensionality reduction to map item embeddings in three domains of the Douban dataset to two-dimensional feature vectors and normalize them onto the unit circle. We show the distribution of normalized features in the unit circle and the distribution of the corresponding angles, for (a) w/o GO (i.e., isolated pre-training without graph coordinators), and (b) HAGO.}
\label{fig:vis}
\end{figure}

Based on this approach, we perform visualizations for the isolated pre-training without graph coordinators (w/o GO), as well as the proposed HAGO, as shown in Figure \ref{fig:vis}.
From the visualization results, it is observable that the item embeddings learned by the model without GO exhibit a significant clustering effect, with the embeddings from the three domains converging in three distinct directions on the unit circle, and their angle distributions are sharper. It indicates that isolated pre-training without graph coordinators fails to capture the connections between nodes from different domains, resulting in a disjointed embedding space across domains. In contrast, for HAGO, it is evident that the feature distributions corresponding to each domain are more uniformly distributed and interweave with one another. This not only suggests that HAGO can capture higher-quality embeddings that are more evenly spread throughout the embedding space, but also that the interweaving distribution of embeddings from different domains indicates HAGO's capability to capture cross-domain associations among items.
This observation underscores the effectiveness of HAGO in enhancing the expressiveness of embeddings by promoting a more cohesive structure within multi-domain graphs, thereby facilitating knowledge transfer across domains.

\subsection{Case Study}\label{app:case}

According to the visualization analysis in Section \ref{sec:visualization}, we observe that the item embeddings learned by HAGO interweave across different domains, as shown in Figure \ref{fig:vis}(b). 
We aim to delve into the specific cross-domain correlations learned by HAGO through case study analysis. 
Our discussion primarily revolves around the Douban-Movie and Douban-Book domains of the Douban dataset, as the semantic relevance between items in these two domains is more intuitive and easily understood by humans.

\begin{figure}[t]
    \centering
    \includegraphics[width=\linewidth]{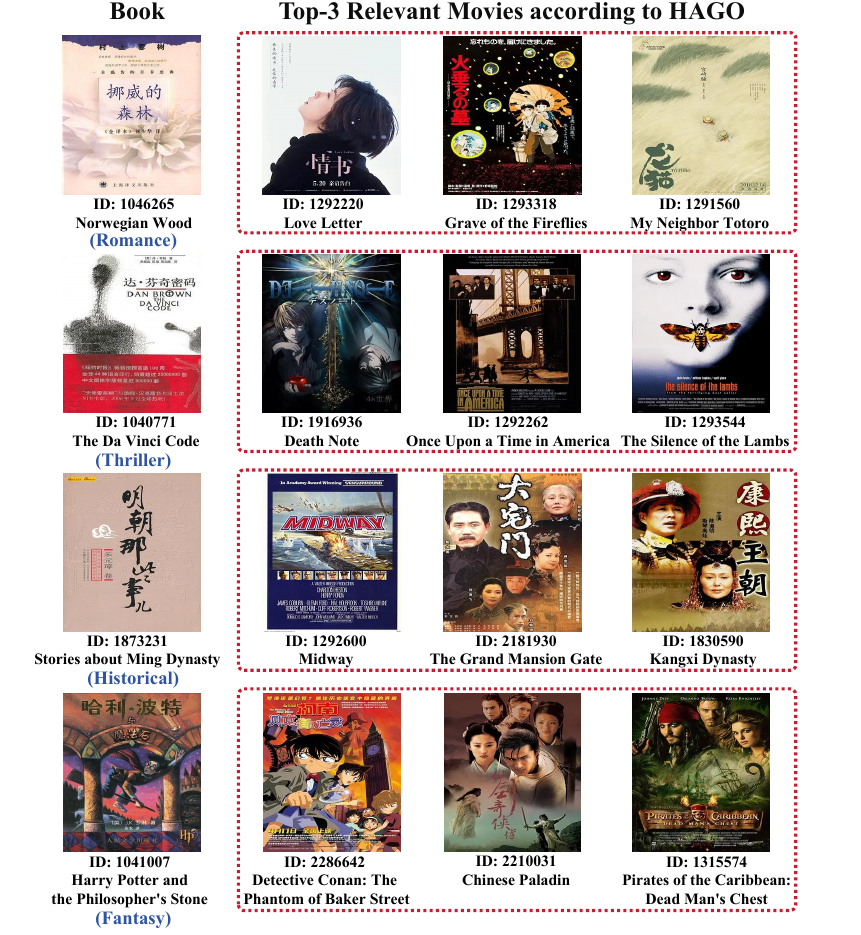}    \caption{Case Study: The most popular books under four specific themes in the Douban dataset, along with top-3 relevant movies for each book thought by HAGO.}
    \label{fig:case}
\end{figure}

Specifically, we select \textbf{the most popular books under four specific themes} (including \textit{romance, thriller, historical, and fantasy}) that appear in the Douban-Book domain. 
For each selected book, we calculate its cosine similarity with all movies in the Douban-Movie domain based on the item embeddings learned by HAGO, and select the corresponding top-3 relevant movies. 
The resulting most popular books and their corresponding top-3 relevant movies for each book are presented in Figure \ref{fig:case}.

For a fair comparison with other SOTA baseline methods, we employ the most typical \textbf{ID-based recommendation paradigm} (i.e., \textit{only use user and item IDs}), without using other features such as item names and tags beyond our research focus. 
However, as observed from Figure \ref{fig:case}, HAGO captures many intriguing associations \textit{without the use of these semantic features}:
For instance, the movies most closely associated with the romance novel ``Norwegian Wood" are all Japanese films with emotional themes (like love, family, and friendship). In contrast, those tied to the thriller novel ``The Da Vinci Code" are primarily crime or mystery genres. Moreover,  users who enjoy the historical novel ``Stories about Ming Dynasty" may also like some movies talking about World War II or Chinese historical stories, while movies associated with ``Harry Potter and the Philosopher's Stone" are all children's favorites such as animations, Chinese Xianxia, and Disney adventures.

It demonstrates that cross-domain knowledge captured through multi-domain collaborative modeling of user-item interaction patterns in HAGO aligns with real semantic relevance.
\section{Conclusions}

Our work delves into the multi-domain representation alignment and the challenging issue of negative transfer in cross-domain recommendation tasks. To tackle these intricate challenges, we introduce the HAGO framework, designed to integrate multi-domain graphs into a cohesive structure. 
We conduct comprehensive experiments to demonstrate the effectiveness of HAGO and vividly discuss its role in representation alignment and the facilitation of cross-domain transfer through visual analysis and case studies.

\begin{acks}
This work is sponsored by Tencent WeChat Rhino-Bird Focused Research Program. Xiangguo Sun and Hong Cheng are supported by project \#MMT-p2-23 of the Shun Hing Institute of Advanced Engineering, The Chinese University of Hong Kong and by grant from the Research Grants Council of the Hong Kong Special Administrative Region, China (No. CUHK 14217622). Chengzhi Piao is supported by grant from the Research Grants Council of the Hong Kong Special Administrative Region, China (No. 12202324). 
\end{acks}

\bibliographystyle{ACM-Reference-Format}
\clearpage

\balance
\bibliography{ref}

\end{document}